\documentclass[useAMS,usenatbib]{mn2e}

\usepackage{natbib, aas_macros}
\usepackage[dvips]{graphicx}
\usepackage{wrapfig}
\usepackage{color}
\usepackage{amsmath}
\usepackage{amssymb}

\newcommand{\Msun}{M_{\odot}}

\newcommand{\fesc}{f_{\rm esc}}

\newcommand{\Nion}{\dot{N}_{\rm Ion}}

\newcommand{\art}{\rm ART^{2}}
\newcommand{\A}{\rm \AA}
\newcommand{\La}{L_{\rm Ly\alpha}}

\newcommand{\Mpc}{\rm {Mpc}}

\newcommand{\lya}{\rm {Ly{\alpha}}}

\newcommand{\Mh}{M_{h}}
\newcommand{\Msunyr}{\rm M_{\odot}~ yr^{-1}}
\newcommand{\fescion}{f_{\rm esc}^{\rm Ion}}
\newcommand{\fescalpha}{f_{\rm esc}^{\rm \lya}}
\newcommand{\fescuv}{f_{\rm esc}^{\rm UV}}

\newcommand{\ergs}{{\rm ergs~s^{-1}}}
\newcommand{\sfruv}{{\rm SFR^{UV}}}
\newcommand{\sfralpha}{{\rm SFR^{Ly\alpha}}}

\title[Escape of Ly$\alpha$ and continuum photons from star-forming galaxies]
{Escape of Ly$\alpha$ and continuum photons from star-forming galaxies}
\author[Yajima et al]
{Hidenobu Yajima$^{1, 2, 3}$\thanks{E-mail:yajima@roe.ac.uk(HY);}, 
Yuexing Li$^{2, 3}$,
Qirong Zhu$^{2, 3}$,
Tom Abel$^{4}$, 
Caryl Gronwall$^{2, 3}$, \newauthor
Robin Ciardullo$^{2, 3}$
\\
$^{1}$ SUPA\thanks{Scottish Universities Physics Alliance}, 
Institute for Astronomy, University of Edinburgh, Royal Observatory, Edinburgh, EH9 3HJ, UK\\
$^{2}$Department of Astronomy and Astrophysics, Pennsylvania State University,
525 Davey Lab, University Park, PA 16802, USA\\
$^{3}$Institute for Gravitation and the Cosmos, The Pennsylvania State University, University Park, PA 16802\\
$^{4}$Kavli Institute for Particle Astrophysics and Cosmology, SLAC National Accelerator Laboratory, Stanford University, \\
2575 Sand Hill Road, Menlo Park, CA 94025, USA\\
}

\begin{document}

\date{Accepted ?; Received ??; in original form ???}

\pagerange{\pageref{firstpage}--\pageref{lastpage}} \pubyear{2008}

\maketitle

\label{firstpage}

%----------------------------------------------------------------------
%
% Abstract
%
%----------------------------------------------------------------------
\begin{abstract}

A large number of high-redshift galaxies have been discovered via their
narrow-band $\lya$ line or broad-band continuum colors in recent years. The 
nature of the escaping
process of photons from these early galaxies is crucial to understanding
galaxy evolution and the cosmic reionization. Here, we investigate the escape
of  $\lya$,  non-ionizing UV-continuum ($\lambda = 1300 - 1600~\A$ in rest
frame), and ionizing photons ($\lambda < 912~\A$) from galaxies by combining a
cosmological hydrodynamic simulation with three-dimensional multi-wavelength radiative transfer calculations. 
 The galaxies are simulated in a box of
  $5^3~h^{-3}\Mpc^{3}$ with high resolutions using the Aquila initial
  condition which reproduces a Milky Way-like galaxy at redshift z=0.
We find that the escape fraction ($f_{\rm esc}$) of these different photons
shows a complex dependence on redshift and galaxy properties:  $\fescalpha$
and $\fescuv$ appear to evolve with redshift, and they show similar, weak
correlations with galaxy properties such as mass, star formation, metallicity,
and dust content, while $\fescion$ remains roughly constant at $\sim 0.2$ from
$z \sim 0 - 10$, and it does not show clear dependence on galaxy properties.
 $\fescalpha$ correlates more strongly with $\fescuv$ than with
$\fescion$. In addition,  we find a relation between the
  emergent $\lya$ luminosity and the ionizing photon emissivity of Lyman Alpha Emitters (LAEs). By combining
  this relation with the observed luminosity functions of LAEs at different
  redshift, we estimate the contribution from LAEs to the reionization of
  intergalactic medium (IGM). Our result suggests that ionizing photons from LAEs alone are not sufficient to ionize IGM at $z \gtrsim 6$, but they can maintain the ionization of IGM at $z \sim 0 - 5$.

\end{abstract}

%----------------------------------------------------------------------
%
% Keywords
%
%----------------------------------------------------------------------

%----------------------------------------------------------------------
%
% Section 1: Introduction
%
%----------------------------------------------------------------------
\section{Introduction}

Young, star forming galaxies can produce  strong $\lya$ emission via hydrogen recombination in the ionized region \citep{Partridge67}. A large population of $\lya$ emitters, or LAEs, has been detected over a wide redshift range, from $2 \lesssim z \lesssim 8.6$ using ground telescopes
 \citep[e.g.,][]{Hu96, Cowie98, Hu98, Steidel00, Rhoads00, Fynbo01, Hu2002, Fynbo03, Rhoads03, Ouchi03, Kodaira03, Maier03, Hu2004, Dawson04, Malhotra04, Horton04, Taniguchi05, Stern05, Kashikawa06, Shimasaku06, Iye06, Hu2006, Gronwall07, Cuby07, Stark07, Nilsson07, Ouchi08, Willis08, Ota08, Hu2010, Ouchi10, Lehnert10, Shibuya12}, and at $z \lesssim 1$ using UV space telescopes \citep{Kunth98, Kunth03, Hayes05, Hayes07, Atek08, Deharveng08, Cowie10}. Therefore, $\lya$ emission may be a powerful tool to study galaxy evolution. 
 
The intrinsic $\lya$ flux is proportional to star formation rate (SFR) when the ionizing photons are absorbed by neutral hydrogen. However, most $\lya$ photons can experience numerous scattering processes by neutral hydrogen due to its large cross section. As a result, $\lya$ photons can have a large optical depth and can be efficiently absorbed by interstellar dust. Hence, although most star forming galaxies are intrinsically LAEs, some may not show a strong $\lya$ line due to efficient dust absorption. It is therefore crucial to study the escape of $\lya$ photons to understand LAEs.

Observationally, some groups have derived the escape fraction of $\lya$ ($\fescalpha$) from the observed flux ratio between $\lya$ and $\rm H\alpha$ at $z = 0.3$ \citep{Atek09}, and at $z=2.2$ \citep{Hayes10}. They found that $\fescalpha$ decreases with color excess $E \rm (B-V)$, albeit with a large dispersion \citep{Atek09}. Alternatively, $\fescalpha$ can be derived from $\lya$ and UV continuum flux \citep{Hayes11, Blanc11}, and it was suggested that $\fescalpha$ monotonically increases with redshift and can be fitted with simple power-law function \citep{Hayes11}. 

Theoretically, semi-analytical studies \citep[e.g.,][]{Kobayashi10} and cosmological hydrodynamics simulations \citep[e.g.,][]{Nagamine10b, Dayal09, Shimizu11} indicate that an adjustment of $\lya$ flux by a factor of $\fesc$ is needed to fit the observed luminosity function or $\lya$ equivalent width distribution. More recently, by combining cosmological hydrodynamics and three-dimensional $\lya$ radiative transfer (RT) simulations,  it was suggested that the escape fraction of $\lya$ photons decreases with halo mass \citep{Laursen09b, Yajima12c}. 

Moreover, the ionizing photons escaping from LAEs may contribute significantly to cosmic reionization. It is widely believed that the intergalactic medium (IGM) is highly ionized at $z \lesssim 6$ from the Gunn-Peterson troughs \citep{Gunn65} in observed quasar (QSO) spectra \citep[e.g.,][]{Fan06a}, and that it should be partially ionized even at $z \gtrsim 10$ from the Thomson scattering optical depth of the cosmic microwave background radiation \citep{Komatsu09}. The possible ionizing sources are thought to be star-forming galaxies, QSOs, and Population III stars. However, since the observed number density of QSOs drops drastically at $z \gtrsim 4$, the ionizing photons from QSOs alone cannot maintain the ionization of the IGM at high redshift \citep{Bolton07}. On the other hand, since the number density of LAEs does not decrease significantly at $z \gtrsim 3$ \citep{Ouchi08}, and the contribution from LAEs to the cosmic star formation rate density becomes dominant at high  redshift \citep{Ciardullo12}, so high-z LAEs may play an important role in the ionization of IGM. 

However, the estimation of ionizing ability of LAEs in previous work suffered from ambiguities in the escape fraction of $\lya$ and ionizing photons. Moreover, an equivalent width (EW) of larger than $20~\A$ is frequently used as the definition of LAEs \citep[e.g.,][]{Gronwall07}. Hence, it is important to understand the difference of escape fraction between $\lya$ and the non-ionizing UV continuum. As the photon escape depends sensitively on the galaxy properties, the ionization state of the interstellar medium (ISM), and the dust content, it is necessary to investigate the radiative transfer of these different photons in galaxies simultaneously. 

In this work, we study the escaping of $\lya$,  non-ionizing UV-continuum ($\lambda = 1300 - 1600~\A$ in the rest frame, hereafter simply ``UV continuum''), and ionizing photons ($\lambda < 912~\A$) from galaxies by combining cosmological hydrodynamic simulations with three-dimensional multi-wavelength radiative transfer calculations. The paper is organized as follows. We describe the galaxy model in 
\S2, and our multi-wavelength radiative transfer code $\art$ in \S3. In \S4, we present the results of the escape of $\lya$, UV-continuum, and ionizing photons, including their evolution with redshift, dependence on galaxy properties, and the contribution of LAEs to cosmic reionization. We discuss the dependence of $\fesc$ on the clumpiness of the ISM, and the limitations of our calculations in \S5, and summarize our findings in \S6.

%\newpage
%----------------------------------------------------------------------
%
% Section 2:  Model and Method
%
%----------------------------------------------------------------------

\section{Galaxy model}

The cosmological simulation presented here follows the formation and evolution of a Milky Way-like galaxy and its substructures. The simulation has been employed in a series of radiative transfer (RT) calculations by \cite{Yajima12b} and \cite{Yajima12c}, and it will be described in detail in Zhu et al. (in preparation). Here we briefly highlight some important features of the simulation. 

It uses the Aquila initial condition, which is the hydrodynamical version \citep{Wadepuhl11} of the initial condition of the Aquarius Project, the largest ever particle simulation of a Milky Way-sized dark matter halo \citep{Springel08a}. The simulation includes dark matter, gas dynamics, star formation, black hole growth, and feedback processes. It was performed using the parallel, N-body/Smoothed Particle Hydrodynamics (SPH) code GADGET-3, which is an improved version of that described in \cite{Springel01} and \cite{Springel05e}. GADGET implements the entropy-conserving formulation of SPH \citep{Springel02} with adaptive particle smoothing, as in \cite{Hernquist89}. Radiative cooling and heating processes are calculated assuming collisional ionization equilibrium \citep{Katz96, Dave99}. Star formation is modeled in a multi-phase ISM, with a rate that follows the Schmidt-Kennicutt Law (\citealt{Schmidt59, Kennicutt98}). Feedback from supernovae is captured through a multi-phase model of the ISM by an effective equation of state for star-forming gas \citep{Springel03a}. The UV background model of \cite{Haardt96} is used. 

The whole simulation falls in a periodic box of $100~h^{-1} \Mpc$ on each side with a zoom-in region of a size $5\times 5\times 5~h^{-3}\Mpc^{3}$. The spatial resolution  in the zoom-in region is $\sim 250~h^{-1}$ pc in comoving scale. The mass resolution of this zoom-in region is $1.8 \times 10^{6}~ h^{-1} \Msun$ for dark matter particles, $3  \times  10^{5}~ h^{-1} \Msun$ for gas, and  $1.5 \times 10^{5}~ h^{-1} \Msun$ for star particles. The cosmological parameters used in the simulation are $\Omega_{m }= 0.25$, $\Omega_{\Lambda} = 0.75$, $\sigma_{8} = 0.9$ and $h=0.73$, consistent with the five-year results of the WMAP \citep{Komatsu09}. The simulation evolves from $z = 127$ to $z = 0$.

In the simulation, a galaxy is identified as a group using on-the-fly friends-of-friends (FOF) group finding algorithms, which links baryon particles to their nearest dark matter neighbor with a dark matter linking length less than 20\% of their mean spacing. 

We note that while our simulation produces a number of properties similar to
those of the MW, such as the morphology, total mass, and star formation rate,
some properties do not agree well with observations. For example, the stellar
to halo mass ratio is somewhat higher than that from abundance-matching analysis of
\citet{Guo10}. Recently, \citet{Scannapieco12} conducted a comprehensive
comparison of the MW simulations using different numerical techniques and
physical models.   He found that the simulated galaxy properties depend
sensitively on the feedback model, and that no simulation to date has produced a
``perfect" MW. However, we should stress that the main purpose of this work is to study the escape of
UV continuum, $\lya$ and ionizing photon from galaxies over a wide range of
redshifts, not producing an exact MW.

\section{The Radiative Transfer Code $\art$}
\label{sec:rt}

The RT calculations are performed using the 3D Monte Carlo RT code, All-wavelength Radiative Transfer with Adaptive Refinement Tree ($\art$), as described in \cite{Li08} and \cite{Yajima12b}. $\art$ features three essential modules: continuum emission from X-ray to radio, $\lya$ emission from both recombination and collisional excitation, and ionization of neutral hydrogen. The coupling of these three modules, together with an adaptive refinement grid, enables a self-consistent and accurate calculation of the $\lya$ properties, which depend strongly on the UV continuum, ionization structure, and dust content of the object. Moreover, it efficiently produces multi-wavelength properties, such as the spectral energy distribution and images, for direct comparison with multi-band observations. The detailed implementations of the $\art$ code are described in \cite{Li08} and \cite{Yajima12b}. Here we focus on the $\lya$ calculations and briefly outline the process. 

The $\lya$ emission is generated by two major mechanisms: recombination of ionizing photons and collisional excitation of hydrogen gas. In the recombination process, we consider ionization of neutral hydrogen by ionizing radiation from stars, active galactic nucleus (AGN), and UV background (UVB), as well as by collisions with high-temperature gas. The ionized hydrogen atoms then recombine and create $\lya$ photons via the state transition $\rm 2P \rightarrow 1S$. The $\lya$ emissivity from the recombination is
\begin{equation}
\epsilon^{\rm rec}_{\alpha} = f_{\alpha } \alpha_{\rm B} h \nu_{\rm \alpha} n_{\rm e} n_{\rm HII},
\end{equation}
where $\alpha_{\rm B}$ is the case B recombination coefficient, and $f_{\alpha}$ is the average number of $\lya$ photons produced per case B recombination. Here we use $\alpha_{\rm B}$ derived in \citet{Hui97}. Since the temperature dependence of $f_{\alpha}$ is not strong, $f_{\alpha} = 0.68$ is assumed everywhere \citep{Osterbrock06}. The product $h\nu_{\alpha}$ is the energy of a $\lya$ photon, 10.2~eV.
 
In the process of collisional excitation, high temperature electrons can excite the quantum state of hydrogen gas by the collision. Due to the large Einstein A coefficient, the hydrogen gas can occur de-excitation with the $\lya$ emission. The $\lya$ emissivity by the collisional excitation is estimated by
\begin{equation}
\epsilon^{\rm coll}_{\alpha} = C_{\rm Ly \alpha} n_{\rm e} n_{\rm HI},
\end{equation}
where $C_{\rm Ly \alpha}$ is the collisional excitation coefficient,
$C_{\rm Ly\alpha} = 3.7 \times 10^{-17} {\rm exp}(- h\nu_{\alpha}/kT) T^{-1/2}~\rm ergs\; s^{-1}\; cm^{3}$ \citep{Osterbrock06}.

Once the ionization structure has been determined, we estimate the intrinsic $\lya$ emissivity in each cell by the sum of above $\lya$ emissivity, $\epsilon_{\alpha} = \epsilon^{\rm rec}_{\alpha} + \epsilon^{\rm coll}_{\alpha}$. 

In the RT calculations, dust extinction from the ISM is included. The dust content is estimated according to the gas content and metallicity in each cell, which are taken from the hydrodynamic simulation. The dust-to-gas ratio of the MW is used where the metallicity is of Solar abundance, and it is linearly interpolated for other metallicity. 
We adopt the dust size distribution of \citet{Todini01} for solar metallicity and a $M = 22~\Msun$ SN model, as in Figure $5$ in their paper. The size distribution is then combined with the dust absorption and scattering cross section of \citet{Weingartner01} to calculate dust absorption opacity curves.
We use the stellar population synthesis model of GALAXEV \citep{Bruzual03} to produce intrinsic spectral energy distributions (SEDs) of stars for a grid of metallicity and age, and we use a simple, broken power law for the AGN \citep{Li08}. A \citet{Salpeter55} initial mass function 
over the range of $0.1 - 100~\rm \Msun$
is used in our calculations. 

In this work, we apply $\art$ to the selected galaxies from the cosmological simulation. 
We pick $\sim 60$ massive galaxies from each snapshot, which gives a total of
936 galaxies. 
The smallest galaxy in the sample has a mass of $\sim 5 \times 10^{8}~\rm \Msun$ at $z=10.4$, 
and it consists of total $\sim 3000$ particles. 
In our post-processing procedure, we first calculate the RT of ionizing photons ($\lambda \le 912 \;\rm \AA$) and estimate the ionization fraction of the ISM. The resulting ionization structure is then used to run the $\lya$ RT to derive the emissivity, followed by the calculation of non-ionizing continuum photons ($\lambda > 912 \;\rm \AA$) in each cell. Our fiducial run is done with $N_{\rm ph} = 10^{5}$ photon packets for each ionizing, $\lya$, and non-ionizing component, which was demonstrated to show good convergence \citep{Yajima12b, Yajima12c}. The highest refinement of the adaptive grid corresponds to a cell size comparable to the spatial resolution of 250 pc in comoving coordinate of the hydrodynamic simulation.
The physical values in each cell are interpolated from neighbor gas particles with the weight of the Kernel function.
The mean uniform density in each cell is used for $\art$ calculations in our models.

%\newpage
%%%%%%%%%%%%%%%%%%%%%%%%%%%%%%%%%%%%%%%

%----------------------------------------------------------------------
%
% Section 3:  Results
%
%----------------------------------------------------------------------

\section{Results}

\subsection{Evolution of the escape fraction of $\lya$ and continuum photons}

\begin{figure}
\begin{center}
\includegraphics[scale=0.43]{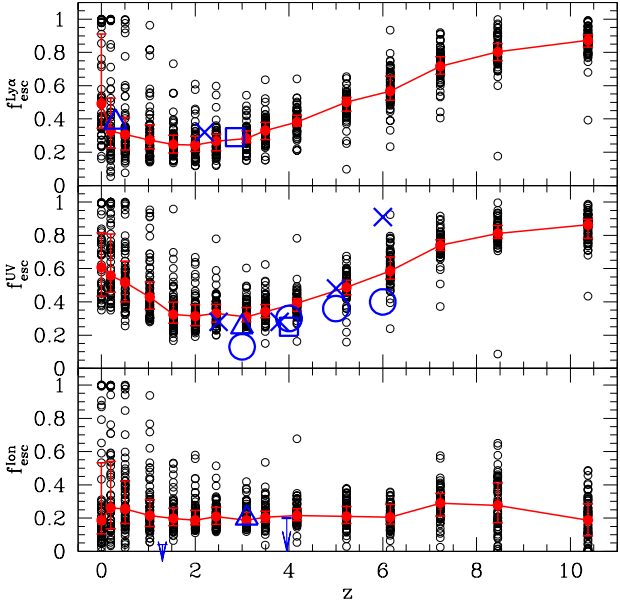}
\caption{
Escape fraction of $\lya$ (upper panel), UV continuum at $1300 \le \lambda \le 1600~\A$ in rest frame (middle panel), 
and ionizing photons (lower panel) as a function of redshift, in comparison with estimates from observations. The black open circles represent individual galaxies from the simulated sample, red filled circles are the corresponding median values at each redshift, with the error bars showing the quartiles. The blue symbols and arrows represent the escape fraction derived from observations. The triangle, cross and square symbols in the upper panel are from \citet{Atek09}, \citet{Hayes10} and \citet{Blanc11}, respectively. The triangle, square, cross and circle symbols in the middle panel are from  \citet{Adelberger00}, \citet{Ouchi04}, \citet{Bouwens09} and \citet{Hathi08}, respectively, which are derived by combining the observed $E \rm (B-V)$ with Calzetti's extinction law \citep{Calzetti00}. The triangle in the lower panel is from \citet{Iwata09}, and the upper limits at $z=1.3$ and $z \sim 4$ come from \citet{Siana10} and \citet{Vanzella10} respectively.
}
\label{fig:fesc_z}
\end{center}
\end{figure}

\begin{figure}
\begin{center}
\includegraphics[scale=0.43]{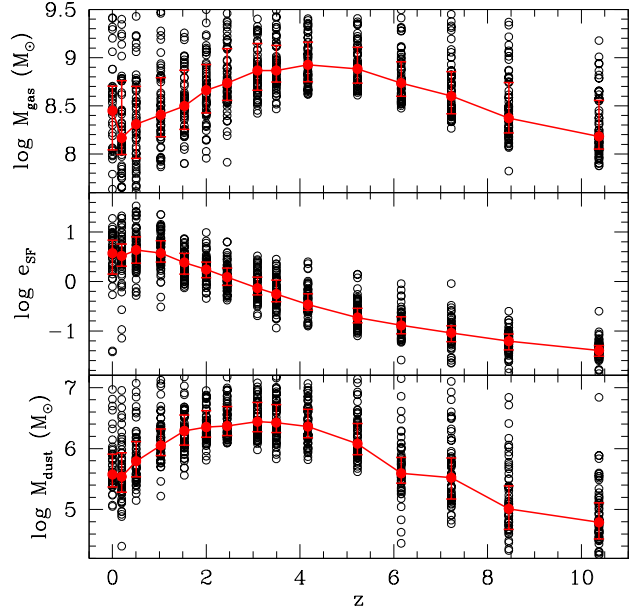}
\caption{
Evolution of various aspects in the simulations. 
%{\it Upper panel:} Mass fraction of gas to dark matter normalized by WMAP-5 year data.
 {\it Upper panel:} Gas mass of each galaxy. 
{\it Middle panel:} Star formation efficiency, defined as $e_{\rm SF} = M_{\rm star} / M_{\rm gas}$. 
{\it Lower panel:} Dust mass. 
The red filled circles correspond to the median values at each redshift, with the error bars showing the quartiles.
}
\label{fig:fgas_z}
\end{center}
\end{figure}

\begin{figure}
\begin{center}
\includegraphics[scale=0.43]{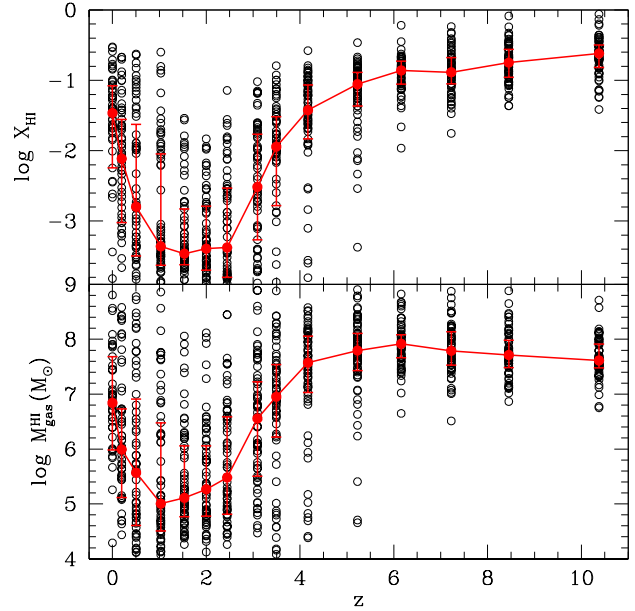}
\caption{
{\it Upper panel:} Mass-weighted mean-values of neutral fraction of hydrogen as a function of redshift.
{\it Lower panel:} Total neutral hydrogen mass of each galaxy.
The red filled circles represent the corresponding median values at each redshift, with the error bars showing the quartiles.
}
\label{fig:xh1}
\end{center}
\end{figure}

\begin{figure}
\begin{center}
\includegraphics[scale=0.5]{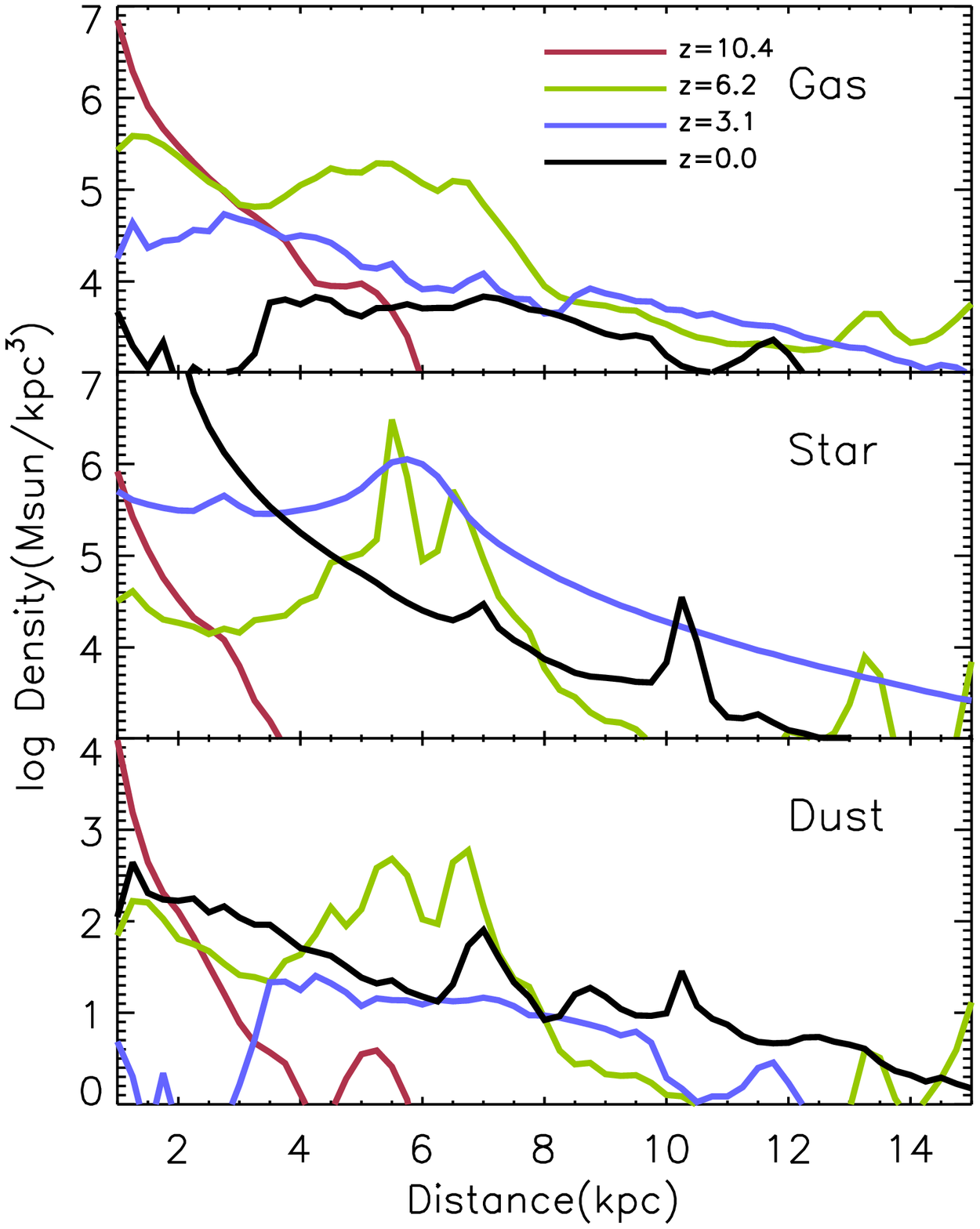}
\caption{
Density profile of the gas (upper panel), stars (middle panel) and dust (lower
panel) of the most massive galaxy from selected redshift as a function of
physical distance from its galactic center (the center of the mass of dark matter halos).
The different color indicates different redshift.
}
\label{fig:density}
\end{center}
\end{figure}

The time evolution of the escape fraction of  $\lya$, UV continuum at $1300 \le \lambda \le 1600~\A$ in rest frame, 
and ionizing photons is shown in Figure~\ref{fig:fesc_z}. 
In this work, we define escape fraction as the ratio of total escaped photons along various angles to total emitted photons, 
i.e., angular mean values. 
We find that $\fesc$ of $\lya$ ($\fescalpha$) moderately evolves with redshift,
and the median values range from $\fescalpha \sim 0.2$ to $0.9$.  $\fescalpha$ somewhat decreases with increasing redshift at $z \lesssim 3$, and then increases with redshift $z > 3$. At lower redshift, since the amount of the gas and dust decreases due to star formation, 
so some fraction of the $\lya$ photons can escape. 

The median $\fescalpha$ remains nearly constant, $\sim 0.3$ at $2 \lesssim z \lesssim 4$, which is similar to the observed value and trend with redshift in LAEs from the HETDEX pilot survey \citep{Blanc11}. The large dispersion in $\fescalpha$, which is similar to the observations \citep[e.g.,][]{Atek09, Hayes10, Blanc11}, may be caused by different galaxy properties such as mass, metallicity and star formation, as we will discuss later in the paper. At $z > 3$, our result is in broad agreement with the implication of \cite{Hayes11}, although the escape fractions of our sample increase  more slowly with redshift. 

At $z \lesssim 3$, however, our results differ from the empirical model of \citet{Hayes11}, in which $\fescalpha$ is parameterized as a power-law function of redshift, $\fescalpha \propto (1+z)^{k}$, with the best fitting power-law index $k = 2.6$. This simple relation implicitly comes from the redshift evolution of metallicity in galaxies suggested by some theoretical predictions \citep[e.g,][]{Kobayashi07} and observations \citep[e.g.,][]{Prochaska03}. 
Dust content can increase with metal in gas, and can effectively absorb $\lya$ photons. 
On the other hand, the distribution of gas and dust can strongly affect  $\fescalpha$ (e.g., Yajima et al. 2012c). 
%At low redshift, the mean gas density decreases,  and a large fraction of gas is consumed by star formation, so $\fescalpha$ of our model galaxies does not decreases with redshift at $z \lesssim 3$, despite the higher metallicity compared to that in higher redshift. 
In addition, at low redshift, a large fraction of dust is consumed by star formation. %as seen in Figure~\ref{fig:fgas_z}.
Figure~\ref{fig:fgas_z} shows the redshift evolution of gas mass, %%gas fraction, 
star formation efficiency, and dust mass.
The dust mass decreases with redshift at $z \lesssim 3$. 
Hence $\fescalpha$ of our model galaxies does not decrease with redshift at $z \lesssim 3$. %despite the higher metallicity compared to higher redshift galaxies.
%In observations, 
Moreover, there is a large dispersion in the observed estimation of $\fescalpha$ and the intrinsic $\lya$ emission, and there is currently no constraint on $\fescalpha$ at $z \sim 0.3 - 2$. Our results are consistent with the observations within the uncertainties.
%Figure~\ref{fig:fgas_z} shows the redshift evolution of \textcolor{red}{gas mass}, %%gas fraction, 
%star formation efficiency, and dust mass.
%The gas fraction is the mass ratio of gas to dark matter of each galaxies normalized by the mass fraction of WMAP-5 data. 
%At $z \lesssim 4$, the gas fraction monotonically decreases with redshift. The median value becomes $\sim 0.16$ at $z \sim 0$. 
At $z \gtrsim 3$, the gas mass increases as redshift decreases because of galaxy growth via gas accretion and merging processes. 
At $z \lesssim 3$, star formation consumes a large fraction of gas and causes a decrease of the gas mass with redshift.
On the other hand, the star formation efficiency, $e_{\rm SF} \equiv M_{\rm star}/M_{\rm gas}$, monotonically increases with decreasing redshift. 
At $z \sim 3$, the stellar mass starts to be larger than the gas mass. 
On the other hand, the evolution of dust mass is not as straightforward.
Until $z \sim 3$, the dust mass increases with redshift due to enrichment by Type-II supernovae, 
however, thereafter it moderately decreases because the dusty gas is consumed by star formation. 
Thus, at lower redshift $z \lesssim 3$, although metallicity is high, $\fescalpha$ does not decrease due to the smaller dust mass. %%and gas mass
%which lead to lower number of scattering and dust absorption.
The gas fraction and star formation efficiency ($e_{\rm SF} \equiv M_{\rm star}/M_{\rm gas}$) can change depending on the feedback models in the simulations. Strong feedback suppress the consumption of gas by star formation, resulting in lower $e_{\rm SF}$. In such a situation, at lower redshift, galaxies may keep having significant gas and dust, and show asmaller escape fraction.
 Our current feedback model can reproduce cosmic star formation rate history over wide range of redshift well while it tends to make too many stars at lower redshift (Zhu et al. in prep.).
 However, the correct prescription of feedback is still under the debate, there is no perfect model to reproduce all observational properties of galaxies. 
 An improved prescription of feedback and star formation will be addressed by a large number of cosmological simulations and comparison with observation in future work (Zhu et al. in prep.).

The evolution of the escape fraction of UV continuum at $\lambda_{\rm rest} = 1300 - 1600~\rm \A$, $\fescuv$, is shown in the middle panel of  Figure~\ref{fig:fesc_z}. Not surprisingly, the general trend is similar to that of $\fescalpha$, as $\fescuv$ is also determined by the dust content and its distribution. However, since the path length of $\lya$ photons before escaping can be longer than that of the UV continuum photons, so the absolute value of $\fescalpha$ is expected to be smaller than that of $\fescuv$. For comparison with observations, this figure also shows $\fescuv$ derived by using observed $E(\rm{B-V})$ with Calzetti's extinction law \citep{Calzetti00}. The attenuation by dust at $1600~\A$ is estimated by $A_{1600} = 10 \times E (\rm{B - V})$ \citep{Ouchi04}. Our result agrees well with the observations, which shows a moderate increase with redshift from $z \sim 3$ to $z \sim 6$. 

On the other hand, the escape fraction of the ionizing photons, $\fescion$, shows a different evolution, as shown in the lower panel of Figure~\ref{fig:fesc_z}. Unlike $\fescalpha$ and $\fescuv$, $\fescion$ does not change significantly with redshift, possibly due to absorption of the ionizing photons by both neutral hydrogen and dust. 
In general, the mean gas density in galaxies increases with redshift, $n_{\rm H} \propto (1+z)^{3}$, and the recombination rate is proportional to $n_{\rm H}^{2}$. Hence at high redshift, it becomes difficult for stellar radiation to ionize the gas. 
Note, however, that this argument is based on a spherical top-hat model, i.e., the mean density of halo is $\sim 18 \pi^2$ times higher than inter-galactic medium (IGM) at when it is virialized \citep{Bryan98}.
In practice, there are many high-density regions due to clumpy structure even at lower redshfit.
In addition, the mass ratio of gas to stars also increases with redshift, therefore, ionizing photons can be more effectively absorbed by interstellar hydrogen gas at higher redshift, leading to a low value of $\fescion$. 
%On the other hand, at low redshift, although the mean gas density decreases, and galaxies are highly ionized by stars and UVB radiation, but the interstellar dust effectively absorbs the ionizing photons. As a result, $\fescion$ does not increase at low redshift. 
On the other hand, at lower redshift $z \lesssim 3$, due to the decrease of SFR \citep[see Figure 2 in][]{Yajima12c}, the neutral fraction of hydrogen increases as shown in the upper panel of Figure~\ref{fig:xh1}.
As a result, H{\sc i} gas mass does not decrease despite the lower total gas mass. This leads to the suppression of $\fescion$ at $z\sim0$. 
Although the H{\sc i} gas mass decreases at  $z \sim 1-2$ due to ionization, 
the increased dust mass contributes to the absorption of ionizing photons.  %, resulting in the suppression of $\fescion$.
Consequently, $\fescion$ shows the weak redshift dependence. 
%since ionizing photons is sensitive to H{\sc i} gas rather than dust unlike $\lya$ and UV photons. 

$\fescion$ is not well constrained observationally, due to a number of uncertainties 
%in the estimation, 
which include the intrinsic SED, the dust attenuation of the UV continuum, and the IGM attenuation. In addition, most observed objects are so faint that only upper limits were obtained on $\fescion$.
$\fescion$ from our simulations is close to the mean observed value at $z \sim 3$ \citep{Iwata09} and the upper limit at $z \sim 4$ \citep{Vanzella10}, but somewhat higher than the upper limit at $z = 1.3$ \citep{Siana10}. This is likely due to the fact that our model galaxies are fainter than the sample of \citet{Siana10}.

Figure~\ref{fig:density} shows the density profiles of gas, stars and dust of the most massive galaxies in the snapshots as a function of physical distance from the galactic center. 
With increasing redshift, gas density becomes high and the distribution is more concentrated towards the galactic center, resulting in the suppression of escape of ionizing photons. On the other hand, the high-density gas induces many scattering of $\lya$ photons, although most of them can successfully escape from galaxies due to the small dust content.
At $z = 0$, a large fraction of gas near the center is converted to stars. 
Near the central region, the gas density decreases with increasing redshift, while the stars increase.
Some bumps in the density profiles are due to the clumpy structure and the merging process. 

Overall, our results of the escape fraction of the $\lya$ and continuum photons are in general agreement with current observations. However, we note that the simulated galaxy sample in this work is very small, and it is limited to progenitors of a MW-like galaxy. We will investigate the $\lya$ and continuum properties of a broader range of galaxy populations from uniform simulations in larger volumes in future work.

\subsection{Relationship between the escape fraction of $\lya$ and continuum photons}

\begin{figure}
\begin{center}
\includegraphics[scale=0.5]{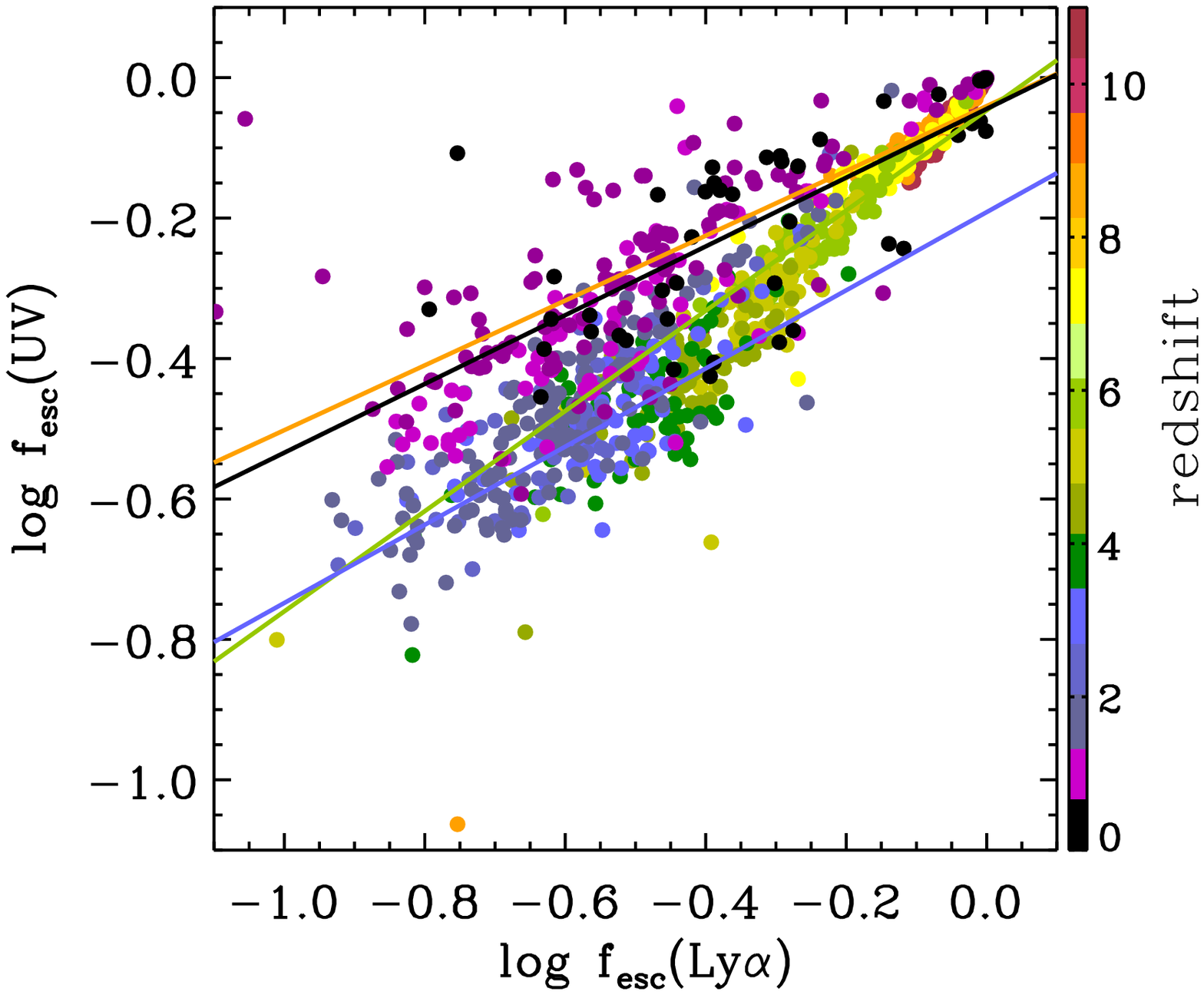}
\includegraphics[scale=0.5]{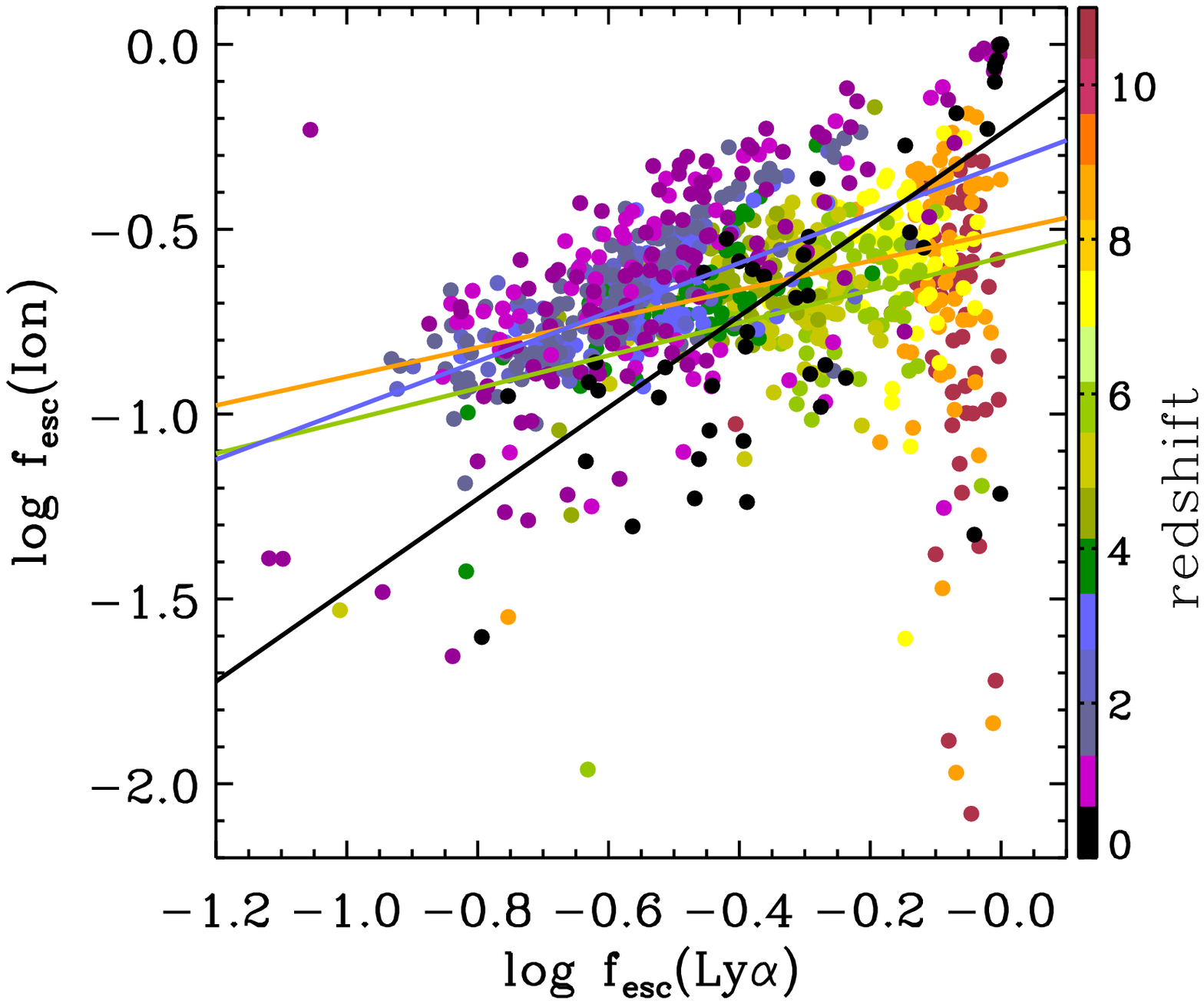}
\caption{The relation of the  escape fractions between $\lya$ and UV continuum at $1300 \le \lambda \le 1600~\A$ in the rest frame ({\it Upper~panel}), and between $\lya$ and ionizing photons ({\it Lower~panel}) at different redshift. Solid lines are absolute-least-deviation fittings with a power-law function, ${\rm log} \fescuv ({\rm or}\: \fescion) = \alpha {\rm log} \fescalpha + \beta$. The color of the points and lines corresponds to different redshifts as specified in the color bar.
}
\label{fig:fesc_uv_a}
\end{center}
\end{figure}

\begin{figure}
\begin{center}
\includegraphics[scale=0.43]{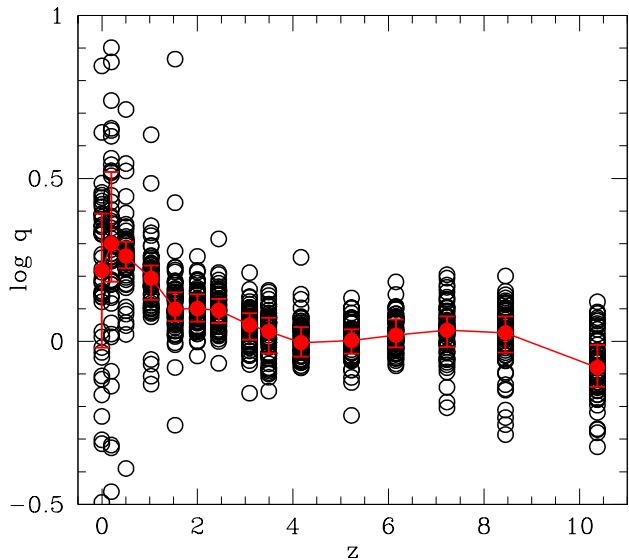}
\caption{
The evolution of  the ratio of optical depths between $\lya$ and UV continuum ($q \equiv \tau_{\lya} / \tau_{UV}$) with redshift. The black open circles represent individual galaxies from the simulated sample, red filled circles are the corresponding median values at each redshift, with the error bars showing the quartiles. 
}
\label{fig:q_z}
\end{center}
\end{figure}

In order to understand the different escape processes of $\lya$ and continuum photons, we study in detail the relation among the escape fractions of $\fescalpha$, $\fescuv$, and $\fescion$ in Figure~\ref{fig:fesc_uv_a}. We assess the strength of the correlation by fitting a simple power-law function, for instance, ${\rm log} \fescuv = \alpha {\rm log} \fescalpha + \beta$, for galaxies at each redshift. The fitting is done on the mean value of each data point, with a bin size of 0.25 dex for the $\fescalpha$.

As shown in the figure, the scatter in the $\fescuv - \fescalpha$ relation (upper panel) appears to be smaller than that of the $\fescion$ - $\fescalpha$ (lower panel). The slope of the former relation falls in the range of $\sim 0.5 - 0.8$ and does not evolve with redshift, while that of the latter decreases with redshift from $\alpha = 1.2$ at $z=0$ to $\alpha=0.1$ at $z=7.2$. This result suggests that $\fescalpha$ is more closely correlated with $\fescuv$ than with $\fescion$. The cause of the different relations is the difference in the escape processes. As mentioned in the previous section,  both $\fescuv$ and $\fescalpha$ depend strongly on the dust content as it is the main absorber of $\lya$ and UV photons, but ionizing photons can also be absorbed by neutral hydrogen gas, so $\fescion$ depends on not only on the dust, but the gas as well. 
In particular, the difference between $\fescalpha$ and $\fescion$ becomes large at higher redshift. 
At high redshift, most of the $\lya$ photons can escape due to the small dust content, while
ionizing photons can be absorbed by nearby clumps of HI gas. 
If star clusters are embedded in the high-density clumps, most of  the ionizing photons are absorbed and $\fescion$ can be very small $\lesssim 0.01$ \citep[see also][]{Yajima11, Kim12}, resulting in the large difference. 

To further demonstrate the escape processes of the $\lya$ and UV continuum and their relation to each other, it is useful to examine the ``$q$'' value, the relative strength of optical depth between $\lya$ and UV continuum, $q \equiv \tau_{\rm \lya} / \tau_{\rm UV}$. Naively, one may think that since the traveling paths of $\lya$ photons are much longer than UV continuum photons due to numerous scattering processes, $\tau_{Lya}$ would be larger than $ \tau_{UV}$, resulting in high $q$ values ($\gg 1$). However, recent observations suggest that $q \sim 1 - 2$ at $z = 0.3 - 3$ \citep{Atek09, Hayes10, Kornei10, Blanc11}. Figure~\ref{fig:q_z} shows the resulting ``$q$'' value from our calculations at different redshifts. It has a mild variation over a wide redshift range $z=0 - 10$, with a median value $q \sim 1 - 2$. This is in good agreement with observations, suggesting that photon escape is a complicated process affected by many factors other than scattering. At high redshift, the dust content is low, and it distributes compactly around galaxy center. Hence, although $\lya$ photons can experience many scattering processes over an extended region, they can escape without dust absorption, leading to $q \sim 1$. At low redshift, since the amount of dust and metallicity increase, $\lya$ photons are more effectively absorbed by dust than the UV continuum, resulting in  a somewhat higher $q$.

\subsection{Dependence of $\fesc$ on galaxy properties}

\begin{figure*}
\begin{center}
\includegraphics[scale=0.5]{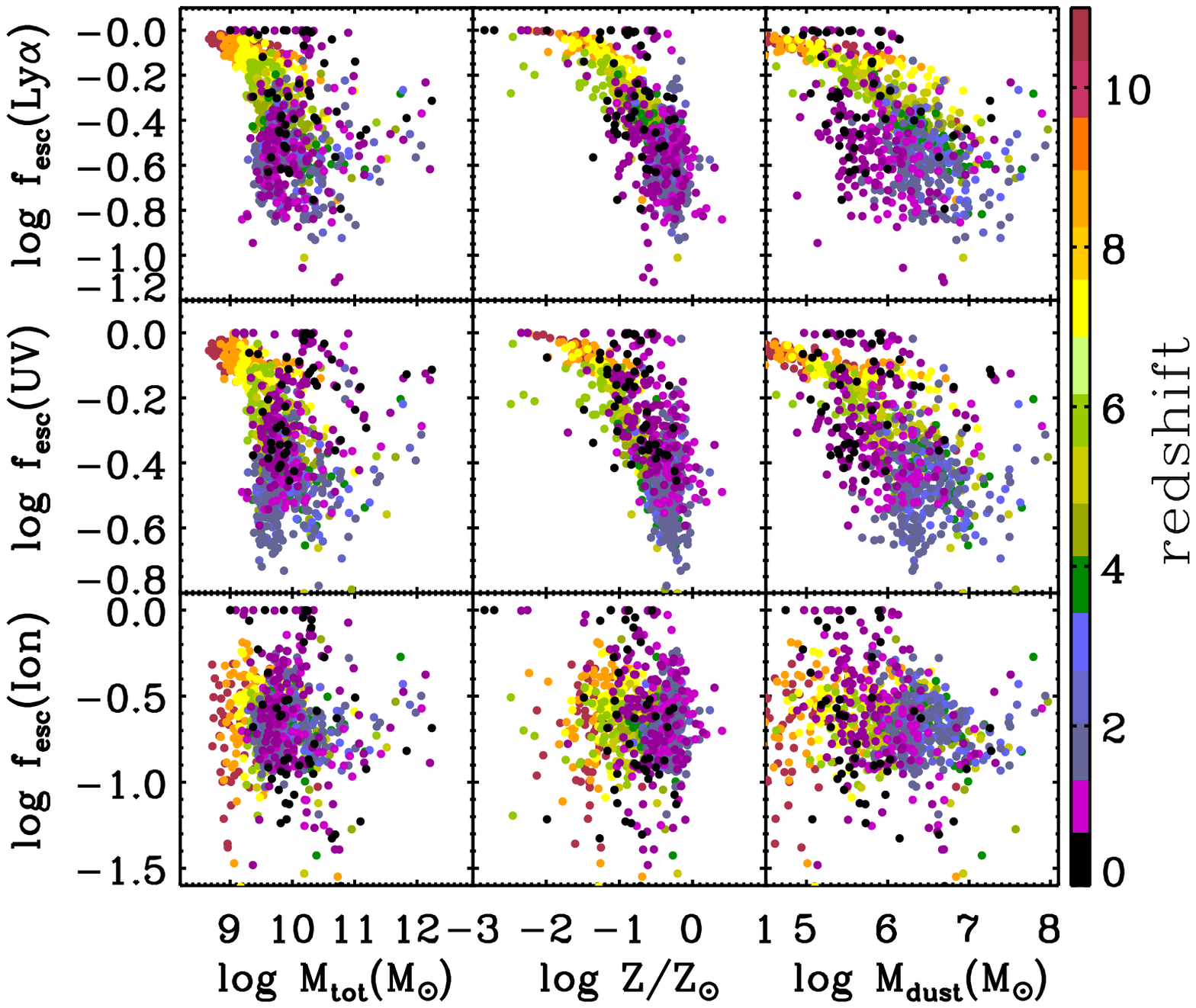}
\includegraphics[scale=0.5]{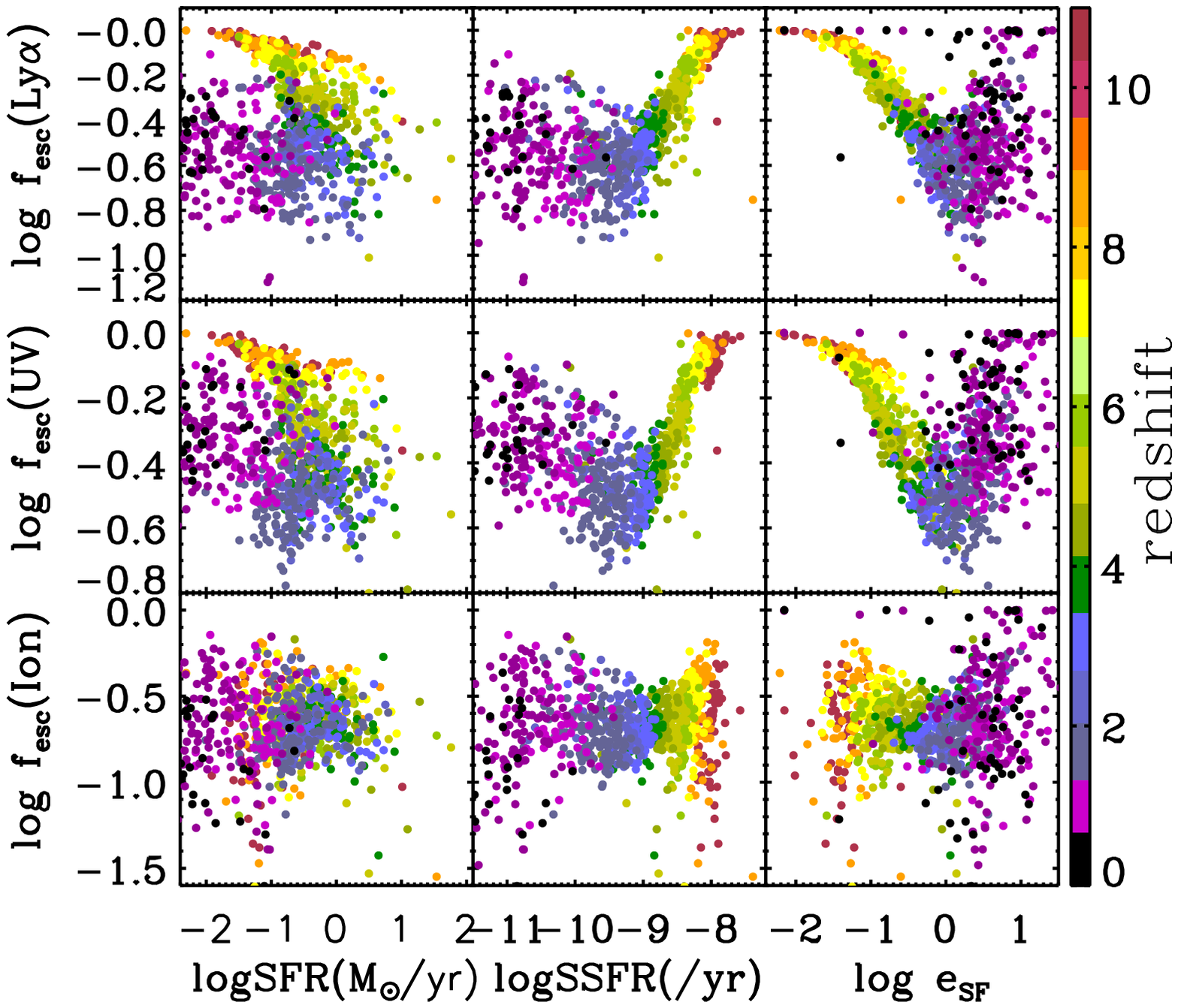}
\caption{Relationship between photon escape fraction, $\fescalpha$ (upper panels), $\fescuv$ (middle panels), and $\fescion$ (lower panels), and galaxy properties, including total galaxy mass ($\rm{M_{tot}}$), gas metallicity $Z$, dust mass ($\rm {M_{dust}}$), star formation rate (SFR), specific SFR (SSFR), and star formation efficiency ($e_{\rm SF}$) defined as the mass fraction of stars to gas.
The color indicates different redshift as shown in the color bar.
}
\label{fig:fesc_mass}
\end{center}
\end{figure*}

The relationships between photon escaping processes and galaxy properties and their evolution are not well known. Here we explore the dependence of photon escape fraction ($\fescalpha$, $\fescuv$, and $\fescion$) on various physical properties of the galaxies from our simulations. As shown in Figure~\ref{fig:fesc_mass}, $\fescalpha$ and $\fescuv$ appear to have similar relationships with galaxy properties such as total mass, metallicity, dust mass, star formation rate (SFR), specific SFR (SSFR), and star formation efficiency. There is no strong correlation between photon escaping and the physical properties of the galaxy, although $\fescalpha$ and $\fescuv$ show a weak declining trend with the galaxy mass, metallicity and dust mass in the interstellar medium, albeit with large dispersions. Such a trend is in broad agreement with observations that metal enriched galaxies tend to have lower $\fescalpha$ \citep{Atek09}. 

In galaxies with high SFRs, there is a large supply of dust and neutral hydrogen gas for star formation, which can lead to efficient scattering and absorption of $\lya$ photons. This trend is in good agreement with the recent semi-analytic work of \cite{Garel12}, which shows suppression of $\lya$ escape by dust in galaxies with high SFRs. In addition,  $\lya$ photons can frequently encounter dust in high-metallicity regions. Therefore, $\lya$ in galaxies with high SFRs, metallicity, and dust mass can be effectively absorbed by dust, resulting in a low $\fescalpha$. Moreover, $\fescalpha$ appears to roughly decrease with halo mass. For example, at $z = 3$, $\fescalpha \sim 0.32$ at $\Mh \sim 10^{10}~\Msun$ to $\sim 0.25$ at $\Mh \sim 10^{11}~\Msun$. This is in good agreement with previous simulations by \citet{Laursen09b}, in which they calculated $\fescalpha$ of nine galaxies at $z = 3.6$ and found a similar relationship. 

However, $\fescion$, does not show any clear dependence on any of these properties. This again may be the result of  more complicated radiative processes of the ionizing photons in dust and neutral hydrogen gas, and it may explain the lack of evolution with redshift seen in the previous section.  
We note that some previous studies reported some correlation between $\fescion$ and halo mass \citep{Razoumov10, Yajima09, Yajima11}. 
The discrepancy may be due to the different galaxy population used in these studies. In our simulation, we focus on a MW-like galaxy, and our galaxy sample is mainly the progenitors of such a galaxy. However, in \cite{Yajima11}, their galaxy sample come from a uniform volume, which may represent the general population seen in the local universe. 
Our simulations do not include old massive populations. The most massive galaxy is the main progenitor of the MW-like galaxy at z=0, and 
evolve without frequent major merging processed. The old massive galaxies do not have a significant fraction of young stars and hence show small $\fescion$. 
On the other hand, $\fescion$ of MW progenitors does not become small due to active star formation of $\gtrsim 5 ~\Msunyr$ at $z \gtrsim 2$ \citep{Yajima12b}.
As a result,  $\fescion$ is  $\gtrsim 0.1$ at $M_{\rm tot} \sim 10^{12}~\rm \Msun$.
In addition, $\fescion$ can change depending on the resolution \citep{Yajima11, Paardekooper11}. 
High-resolution simulations can resolve high-density gas clumps around young star clusters in low mass galaxies
and the gas clumps can absorb ionizing photons efficiently, resulting in smaller $\fescion$. 
The resolution of our cosmological simulations is much higher than \citet{Yajima11}, and the RT calculations use 
AMR grid structure while \citet{Yajima11} used uniform grids. 
Hence, our simulations may show smaller $\fescion$ at lower galaxy mass, and no strong correlation with mass.
On the other hand, our simulations show the large dispersion at lower mass, and $\fescion$ at lower mass can be very high $\sim 1.0$
and small $\sim 0$, the trend is similar with \citet{Yajima11}. 
The variation of offset distance between star clusters and high-density clumps causes the large dispersion \citep{Yajima11, Kim12}.
We plan to extend this study to a larger population of galaxies from larger, uniform simulations with higher resolution.

\subsection{Contribution of LAEs to cosmic reionization}

\begin{figure}
\begin{center}
\includegraphics[scale=0.5]{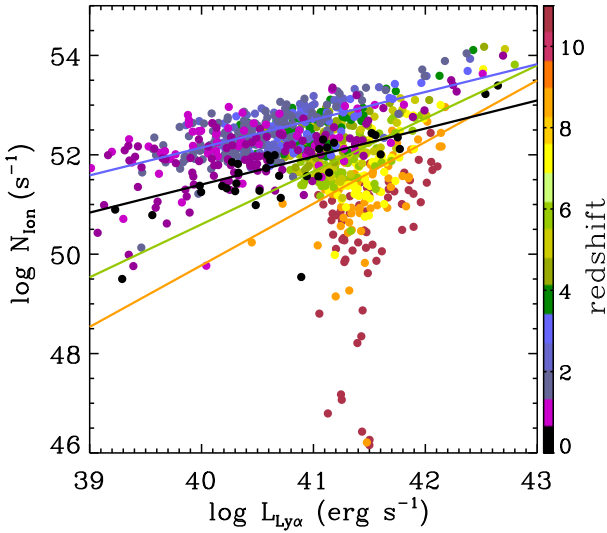}
\caption{
Emissivity of ionizing photons as a function of emergent $\lya$ luminosity.
}
\label{fig:La_Ion}
\end{center}
\end{figure}

\begin{figure}
\begin{center}
\includegraphics[scale=0.4]{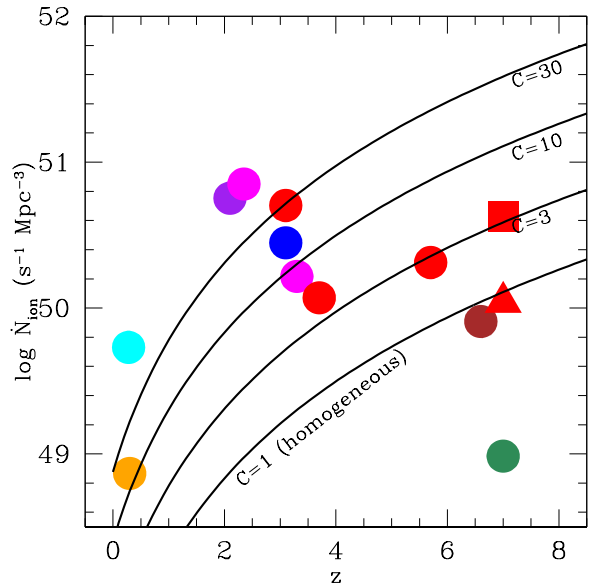}
\caption{
Ionizing photon emissivity of LAEs at different redshifts (colored symbols), in comparison with theoretical predictions of the number of ionizing photons needed for ionizing the IGM (solid lines).  The emissivity is derived from the integration of the $\lya$ luminosity function
 converting from $\La$ to ionizing photon number. Different color circle indicates different luminosity function used: 
\citet[][orange]{Cowie10}, \citet[][cyan]{Deharveng08}, \citet[][purple]{Guaita10}, \citet[][magenta]{Blanc11}, \citet[][blue]{Gronwall07}, \citet[][red]{Ouchi08}, \citet[][brown]{Ouchi10} and \citet[][green]{Hibon12}. 
The filled circles represent the integration of the luminosity function to the detection limit. %%and $\La = 10^{39}~\ergs$, respectively. 
%Red triangle at $z = 7$ is the emissivity by using the luminosity function at $z = 5.7$ in \citet{Ouchi08} \textcolor{blue}{and our result between $\Niion$ and $\La$ at $z=7$}. 
Red square and triangle at $z = 7$ are the emissivity by using the luminosity function at $z = 5.7$ in \citet{Ouchi08} and our result between $\Nion$ and $\La$ at $z=7$
with and without the effect of IGM transmission $T_{\lya} = 0.26$ at $z=5.7$ \citep{Laursen11}, respectively.
The number of ionizing photons needed for ionizing the IGM are calculated based on the model of \citet{Madau99}, $\Nion \;({\rm s^{-1} \: Mpc^{-3}}) = 10^{47.4} C (1+z)^{3}$, where $C$ is the clumpiness of IGM.
}
\label{fig:reion}
\end{center}
\end{figure}

\begin{figure}
\begin{center}
\includegraphics[scale=0.4]{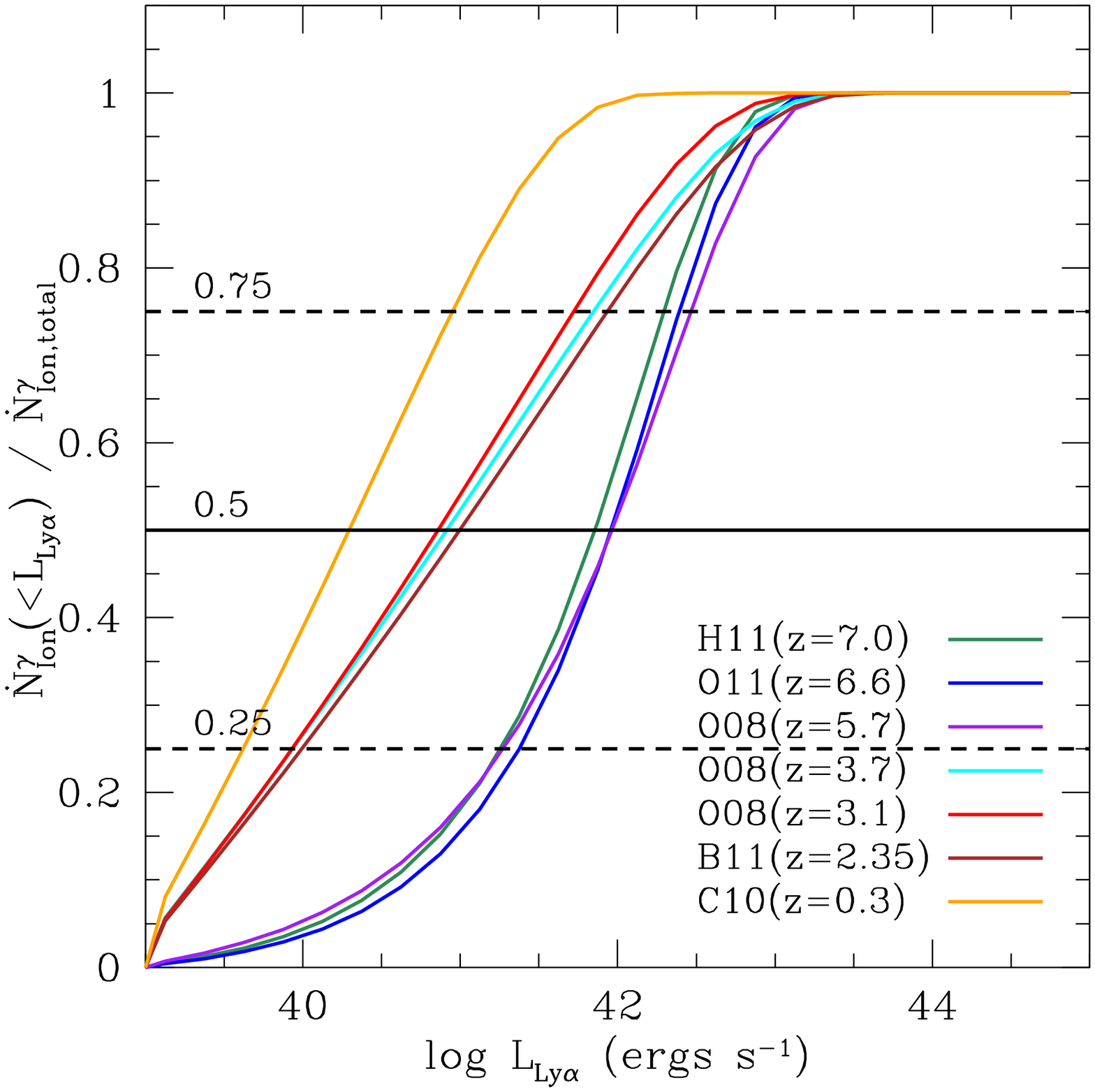}
\caption{
The stacked emissivities of ionizing photons normalized to total emissivity. Different color indicates different observations used in the derivation of ionizing photon emissivity from the $\lya$ luminosity function: \citet[][orange]{Cowie10},  \citet[][brown]{Blanc11},
 \citet[][red : $z=3.1$, cyan : $z = 3.7$ and purple : $z = 5.7$]{Ouchi08},
\citet[][blue]{Ouchi10} and \citet[][green]{Hibon12}.
}
\label{fig:ionfrac}
\end{center}
\end{figure}

The escaping of photons from early galaxies has a significant role in the reionization of neutral hydrogen. However, the estimation of ionizing photon emissivity ($\Nion$) of observed LAEs has been difficult due to the ambiguity of the escape fraction of $\lya$ and ionizing photons. Although the intrinsic emissivity of ionizing photons is roughly related to SFR which can be derived from $\lya$ luminosity,
the estimation of the intrinsic $\lya$ luminosity depends strongly on $\fescalpha$. In addition, the emissivity of ionizing photons of galaxies can be greatly changed by $\fescion$. Therefore, an accurate estimate of $\fescalpha$ and $\fescion$ is critical to understand the cosmic reionization history. Here we directly estimate the $\Nion$ of observed LAEs by using our results of the escape fractions of $\lya$ and ionizing photons.

We at first evaluate the relation between $\Nion$ and emergent $\La$. As shown in Figure~\ref{fig:La_Ion}, although $\Nion$ roughly increases with $\La$, there is a large scatter. We fit the data from each redshift with a power-law function, ${\rm log} \Nion = \alpha {\rm log} \La + \beta$.
The slope $\alpha$ changes with redshift, $\sim 0.7 \;({\rm z=0.0}),~0.8\; ({\rm z=3.1}),~1.3\; (\rm z=6.2)$ and $1.9~ ({\rm z=10.2})$.
At higher redshift, typical galaxy mass decreases, and they have smaller $\fescion$, as seen in Figure~\ref{fig:fesc_z}. Moreover, the $\La$ of these low-mass, high-redshift galaxies can be boosted significantly  by the excitation $\lya$ cooling \citep{Yajima12c}. Therefore, the slopes at higher  redshift have larger values than their lower redshift counterparts. Next, we integrate the luminosity function (LF) of observed LAEs, and by converting $\La$ to $\Nion$ using the above fitting formula, we derive the emissivity of ionizing photons from these LAEs. 
 
Figure~\ref{fig:reion} shows the resulting emissivity at different redshifts, in comparison with theoretical predictions of the number of ionizing photons needed for ionizing the IGM. 
%Filled and open symbols indicate the emissivity derived from the integration of the LF up to the observational limit and {$\La = 10^{39}~\ergs$}, which is fainter than detection limit of current observations by order 2-3 \citep[e.g., $\sim 10^{42}~\ergs$ : ][]{Ciardullo12}. 
Filled symbols indicate the emissivity derived from the integration of the LF up to the observational limit.
Solid lines are the number of ionizing photons needed for ionizing the IGM based on the model of \citet{Madau99}, $\Nion \;({\rm s^{-1} \: Mpc^{-3}}) = 10^{47.4} C (1+z)^{3}$, where $C$ is the clumpiness of IGM. Recent simulations suggest $C \sim 3 - 10$ \citep[e.g.,][]{Iliev07, Pawlik09b}. Note that $C=1$ corresponds to a homogeneous IGM, hence the clumpiness must be larger than unity in the cosmological model based on hierarchical formation.
If the emissivity from LAEs is on or above the solid lines, then it means that LAEs alone can ionize the IGM.

As shown in the figure, the emissivity is larger than the $C=10$ line at $z \sim 0 - 4$, and hence they can maintain the ionization of the IGM. At $z = 5.7$, the emissivity is marginally close to the $C=3$ line. 
At $z \gtrsim 6$, the emissivity is below the $C=1$ line.  %%, and it is below the $C=3$ line even if we integrate the LF down to $\La=0$. 
However, the observed LF at $z \gtrsim 6$ could be suppressed by IGM scattering. Hence, the LF at $z \sim 5.7$ is frequently used as an intrinsic LF at $z > 6$ \citep[e.g.,][]{Kashikawa11}, motivated from no evolution of the observed LFs from $z \sim 3.1$ to $5.7$ \citep{Ouchi08}. Hence, here we estimate the emissivity at $z=7$ by using the LF at $z=5.7$ of \citet{Ouchi08} and the relation between $\La$ and $\Nion$ at $z=7$. However, the derived emissivity is below the $C=3$. % line even considering the upper limit. 
Therefore, we suggest that LAEs alone cannot ionize the IGM at $z \gtrsim 6$. 
On the other hand, even at $z=5.7$, some fraction of $\lya$ flux can be scattered by the  IGM. 
Recent simulations show the transmission rate in the IGM is $f_{\rm IGM} \sim 0.26$ at $z=5.8$ \citep{Laursen11}.
Due to the IGM scattering, we may underestimate the emergent $\lya$ flux of observation at high redshift, 
resulting in an underestimation of  the ionizing photon emissivity. 
Hence, we use the LF at $z=5.7$ with the boost factor of $1/0.26$, and then derive the $\Nion$ at $z=7$ using the relationship between $L_{\lya}$ and $\Nion$ which is shown by the filled square. 
As a result,  the estimate of $\Nion$ becomes $\sim 4$ times higher than that without IGM attenuation,
and marginally reaches  $C=3$. 
Therefore, if the LF at $z > 6$ can be estimated with the accurate IGM transmission,
observed LAEs can be the main sources of cosmic reionization at $z \sim 7$. 
However, the degree of  ionization of the IGM around high-redshift LAEs is still under the debate, resulting in  large uncertainties in the IGM transmission.
In addition, the IGM transmission is sensitive to the shape of line profile, e.g., $\lya$ photons in the red wing are not significantly scattered by the IGM. 
Large-scale cosmological simulations with radiative transfer calculations in the ISM and the IGM are required for deriving the IGM transmission at high redshift. 
We plan to investigate such simulations in future work. 

It has been suggested that other populations (e.g., LBGs, Population III stars, faint quasars) may be needed to ionize the neutral hydrogen at $z \gtrsim 6$. It was shown that the emissivity from LBGs can be larger than LAEs \citep{Yajima09}. However, \citet{Ouchi09b} showed that  the emissivity from LBGs at $z \sim 7$ falls below the $C=1$ line.  Another possibility to ionize the IGM is a correction of the faint-end slope of LF. In observations of LAEs at $z \gtrsim 3$ \citep[e.g.,][]{Ouchi10, Ciardullo12}, the slope is assumed to be $- 1.36 \sim - 1.65$, which is derived from recent deep survey at lower redshift \citep[e.g.,][]{Cassata11}, because observations of high redshifts can only trace the bright-end of the LF. If this slope is steeper (e.g., $\alpha \lesssim -2$), then the contribution from faint galaxies becomes large \citep[e.g.,][]{Bouwens11, Jaacks11}, giving rise to higher emissivity which can ionize the IGM. The next generation telescopes which have higher sensitivity (e.g., JWST, GMT, TMT) will be able to measure the slope more precisely.

Figure~\ref{fig:ionfrac} shows the stacked emissivity as a function of $\La$ from observations of LAEs. 
The stacked emissivity is estimated by integrating from $10^{39}~\ergs$ for including fainter sources than the detection limit of observations. 
The emissivity increases with redshift,
as the slope of fitting function between $\Nion$ and $\La$ becomes steeper at higher redshift. About half of the ionizing photons comes from LAEs of $4.2 \times 10^{41} \le \La \le 4.2 \times 10^{42}~\ergs$ at $z = 6.6$ \citep[LF by ][]{Ouchi10}, $2.4 \times 10^{41} \le \La \le 2.4 \times 10^{42}~\ergs$ at $z = 7$ \citep[LF by ][]{Hibon12}. Hence, typical observed LAEs around $\La^{*}$ can be the main contributors to the IGM ionization among the LAE population. On the other hand, at $z=3$, half of ionizing photons come from LAEs of $1.3 \times 10^{40} \le \La \le 7.5 \times 10^{41}~\ergs$ \citep[LF by ][]{Ouchi08}. Therefore, LAEs fainter than detection limit of most observations \citep[e.g., $\La \sim 10^{42}~\ergs$ : ][]{Ouchi08} may contribute significantly to the ionization of the IGM.
Note that we here apply our simulation results for the observed LAEs in general fields via using the LFs. % in the estimation of $\Nion$.
However, our simulations use  the small box and the specific condition, hence %our galaxy sample does not include massive galaxies and a wide range of galaxy environments.  
 the relation between $\Nion$ and $\La$ may differ from that of general fields. %somewhat change in different environments. 
%To derive the contribution from LAEs to cosmic reionization more accurately, 
To take LAEs of various environments into the account, 
we will investigate a larger sample in a large volume ($\gtrsim 100^{3}~\rm Mpc^{3}$) in our future works.

%%%%%%%%%%Figure

%----------------------------------------------------------------------
%
% Section 4:  Discussion
%
%----------------------------------------------------------------------

\section{DISCUSSION}

\subsection{Star formation rate from emergent $\lya$ and UV flux}

\begin{figure}
\begin{center}
\includegraphics[scale=0.5]{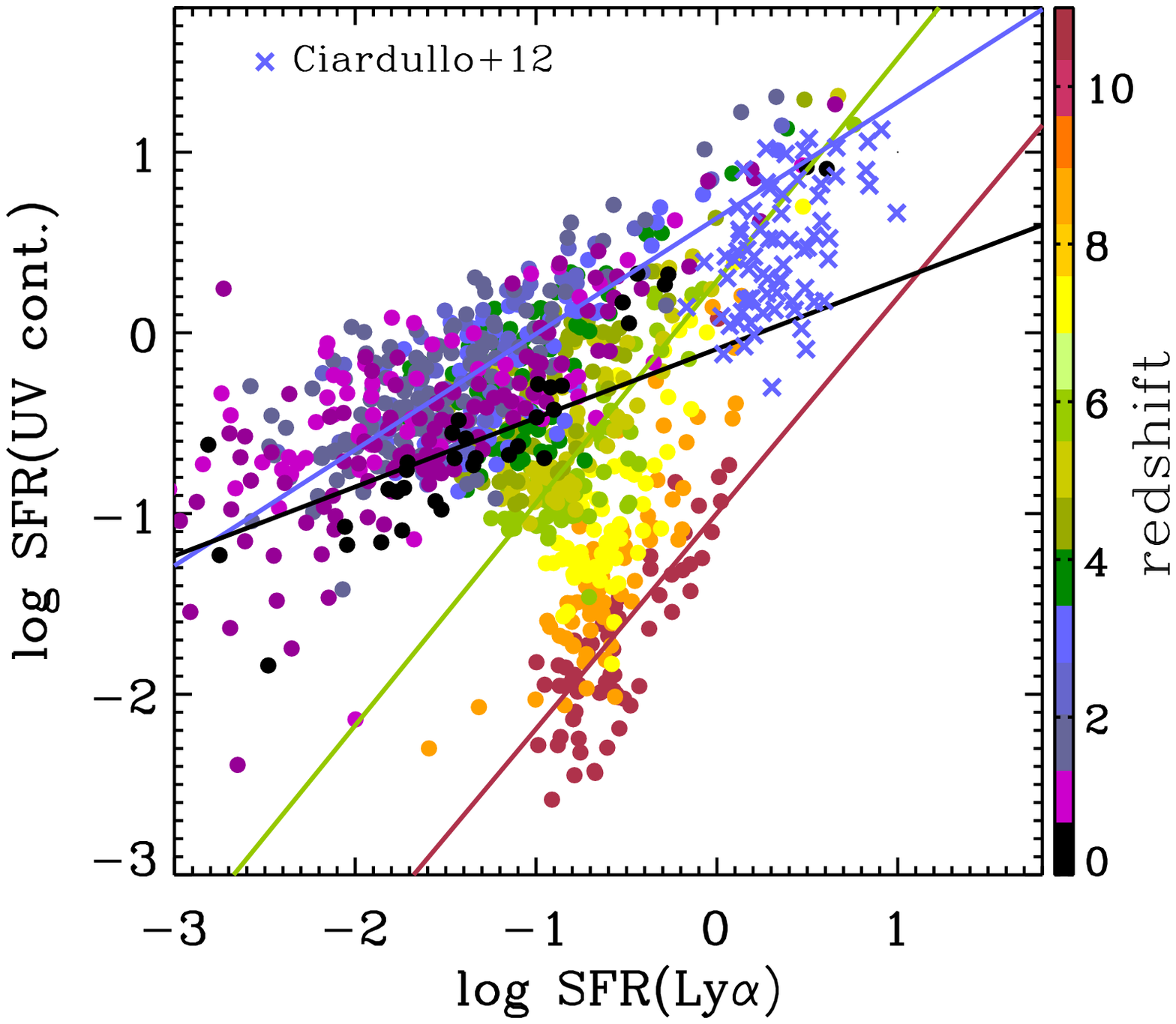}
\caption{
Comparison of SFRs derived from the emergent UV continuum and $\lya$ photons. The estimation of SFR from UV continuum is based on 
Equation (1) in \citet{Kennicutt98}. The SFR from $\lya$ is estimated using equation (2) in \citet{Kennicutt98} with the assumption of case B recombination.
The cross symbols are the derived SFRs of spectroscopically confirmed LAEs at $z =3.1$ from \citet{Ciardullo12}. The color indicates different redshift as shown in the color bar.
}
\label{fig:sfr_comp}
\end{center}
\end{figure}

The observed $\lya$ and UV continuum fluxes are frequently used to derive SFRs. However, this method is subject to the uncertainty of the photon escape fraction, and depends on model
assumptions.  Thus, the derived SFRs may differ by orders of magnitude. To illustrate this point, we show in  Figure~\ref{fig:sfr_comp} the comparison of SFRs derived from $\lya$ and UV continuum flux at different redshifts. 

The SFRs of some bright LAEs at $z=3.1$ from our simulations are consistent with the recent observation of \citet{Ciardullo12}, but the dispersion is large. Since $\fescuv$ is higher than $\fescalpha$, the derived $\sfruv$ is somewhat larger than $\sfralpha$, and close to the true SFR. Most of our model galaxies are fainter than the observational sample, so our results represent the SFR relation of faint LAEs, and suggest  
that it roughly can be fitted by the same power-law relationship.

On the other hand, at $z \gtrsim 6$, $\sfralpha$ becomes larger than $\sfruv$. At this high redshift, the contribution of excitation $\lya$ cooling to the emergent $\lya$ luminosity becomes dominant \citep{Yajima12c}. The $\lya$ cooling due to excitation is no longer directly related to SFR, so  it can result in an overestimate of  the SFR, because the derivation of SFR by $\lya$ flux as in \citet{Kennicutt98} considers only stellar radiation.
Deep multi-band surveys by the next generation telescopes (e.g., GMT, TMT) may trace such a tail of %high SFR by $\lya$ which may be evidence of dominant $\lya$ source by excitation cooling process. 
higher SFR estimated by $\lya$ than by UV continuum, which may be evidence of the strong excitation $\lya$ cooling.
Moreover, the contribution of $\lya$ cooling due to excitation becomes larger in lower-mass galaxies due to lower star formation rates. Since SFR roughly increases with galaxy mass, the gradient between $\sfruv$ and $\sfralpha$ becomes large. If we fit them by a power-law function, the power-law index is $\alpha = 0.6$ ($z = 3.1$), and {$1.2$} ($z = 10.4$)

\subsection{Dependence of ISM clumpiness}

\begin{figure}
\begin{center}
\includegraphics[scale=0.5]{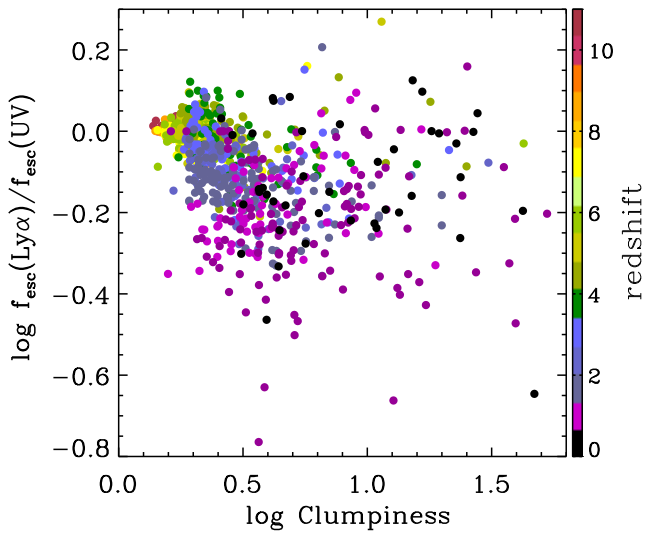}
\caption{
The ratio of escape fraction of $\lya$ to UV continuum photons 
as a function of the clumpiness of the interstellar gas, which is defined by
$<\rho^{2}> / <\rho>^{2}$, where $\rho$ is gas density. The color indicates different redshift as shown in the color bar.
}
\label{fig:fesc_c}
\end{center}
\end{figure}

It was suggested by \citet{Neufeld91} that  large $\lya$ EWs  may be produced by a clumpy interstellar media, because $\lya$ photons can be scattered by neutral hydrogen on the surface of clouds before dust absorption, and they can escape without much dust absorption in the clumpy dusty media. On the other hand, UV continuum photons can be absorbed by dust passing through the clouds. As a result, $\fescalpha$ can be larger than $\fescuv$, causing higher EW. \citet{Hansen06} showed that $\lya$ can effectively escape in the idealized clumpy medium
by RT simulations.

Our simulations provide a unique opportunity to study the effect of ISM clumpiness on the escape of $\lya$ and UV continuum photons. 
Figure~\ref{fig:fesc_c} shows the ratio between $\fescalpha$ and $\fescuv$ as a function of the clumpiness of the interstellar gas, defined as 
$C = <\rho^{2}> / <\rho>^{2}$. We find that  $\fescalpha$ is mostly smaller than  $\fescuv$, and that the ratio does not correlate with  clumpiness. In our simulations, the $\lya$ photons can be created in high-density clouds, because star formation can occur in such a region and partially ionize the gas. In such a situation, the $\lya$ photons are effectively absorbed by dust at the origin place. Hence,  $\fescalpha$ cannot be higher than $\fescuv$ even in a clumpy structure. Our results suggest that there is no correlation between the $\lya$ EW and the clumpiness of the ISM.

%----------------------------------------------------------------------
%
% Section 5:  Summary
%
%----------------------------------------------------------------------

\section{SUMMARY}

In this paper, we have investigated the escape of  $\lya$,  the non-ionizing UV continuum, and ionizing photons from galaxies by combining cosmological SPH simulations and three-dimensional multi-wavelength radiative transfer calculations. We found that the escape process differs significantly from one photon species to another.  $\fescalpha$ correlates more strongly with $\fescuv$ than with $\fescion$. Both $\fescalpha$ and $\fescuv$ have similar evolution trends with redshift, and they show similar, weak dependence on galaxy properties such as galaxy mass, metallicity, dust content, star formation rate, specific star formation rate, and star formation efficiency. However, $\fescion$ does not show  evolution with redshift and it shows no clear correlation with galaxy properties. These different behaviors may be explained by the different escaping mechanisms. Both $\fescuv$ and $\fescalpha$ depend strongly on the dust content as it is the main absorber of $\lya$ and UV photons, but ionizing photons can also be absorbed by neutral hydrogen gas, so $\fescion$ depends on not only the dust, but the gas as well. 

In addition, we estimated the ionizing photon emissivity of LAEs and their contribution to the ionization of the IGM, by combining our simulations with the observed luminosity functions of LAEs at different redshifts. We found that the ionizing photons from LAEs can maintain the ionization of  the IGM at $z \sim 0 - 4$, but they are not sufficient to ionize the  IGM at $z \gtrsim 6$.  Other populations (e.g., LBGs, Pop III stars, faint quasars) may be needed to ionize the neutral hydrogen.  

Moreover, we found that the star formation rates derived from emergent $\lya$ and UV may subject to uncertainty of orders of magnitude depending on the photon escape fractions. This would have significant implications in the interpretation of observational data. Furthermore, we found no correlation between the EW of $\lya$ line and the clumpiness of the ISM as suggested by previous studies. 

Overall, our simulations of the escape fraction of the $\lya$ and continuum photons are in general agreement with current observations. However, we note that the galaxy sample in this work is very small, and it is limited to the progenitors of a MW-like galaxy. We will investigate the properties of $\lya$ and continuum photons of larger galaxy populations from uniform simulations in larger volumes in future work.

%----------------------------------------------------------------------
%
% Acknowledge
%
%----------------------------------------------------------------------
\section*{Acknowledgments}
We thank Mark Dijkstra, Claude-Andr{\'e} Faucher-Gigu{\`e}re, Eric Gawiser, Matt Hayes and Lars Hernquist for stimulating discussions and helpful comments. Support from NSF grants AST-0965694, AST-1009867 (to YL), AST-0807075 (to TA) and AST-0807885 (to CG \& RC)
is gratefully acknowledged. YL thanks the Institute for Theory and Computation (ITC) at Harvard University where the project was started for warm hospitality. We acknowledge the Research Computing and Cyberinfrastructure unit of Information Technology Services at The Pennsylvania State University for providing computational resources and services that have contributed to the research results reported in this paper (URL: http://rcc.its.epsu.edu). The Institute for Gravitation and the Cosmos is supported by the Eberly College of Science and the Office of the Senior Vice President for Research at the Pennsylvania State University.

%----------------------------------------------------------------------
%
% References
%
%----------------------------------------------------------------------
%\begin{thebibliography}{99}

%\bibliographystyle{mn}

%\bibliography{mn-jour,HY}

%\end{thebibliography}

\label{lastpage}

\end{document}